\documentclass[showpacs,twocolumn,amsmath,superscriptaddress]{revtex4-1}
\usepackage[dvipdfmx]{graphicx}
\usepackage{color,amsmath,lmodern,bm,array,braket}

\newcommand{\br}{{\bm r}}
\newcommand{\be}{{\bm e}}
\newcommand{\bn}{{\bm n}}

\newcommand{\cH}{\mathcal{H}}
\newcommand{\cO}{\mathcal{O}}

\newcolumntype{C}[1]{>{\centering\arraybackslash}p{#1}}
\newcolumntype{L}[1]{>{\raggedright\arraybackslash}p{#1}}
\newcolumntype{R}[1]{>{\raggedleft\arraybackslash}p{#1}}

\begin{document}

\preprint{Lichtenstein}

\title{Local force method 
for the {\it ab initio} tight-binding model with spin-dependent hopping
}%

\author{Takuya Nomoto}%
\email{nomoto@ap.t.u-tokyo.ac.jp}%
\affiliation{Department of Applied Physics, The University of Tokyo, Hongo, Bunkyo-ku, Tokyo, 113-8656, Japan}%
\author{Takashi Koretsune}%
\affiliation{Department of Physics, Tohoku University, Miyagi 980-8578, Japan}%
\author{Ryotaro Arita}%
\affiliation{Department of Applied Physics, The University of Tokyo, Hongo, Bunkyo-ku, Tokyo, 113-8656, Japan}
\affiliation{RIKEN Center for Emergent Matter Science (CEMS), Wako 351-0198, Japan}
\date{\today}%

%Abstract
\begin{abstract}
To estimate the Curie temperature of metallic magnets from first principles, we develop a local force method for the tight-binding model having spin-dependent hopping derived from spin density functional theory. While spin-dependent hopping is crucial for the self-consistent mapping to the effective spin model, the numerical cost to treat such non-local terms in the conventional Green's function scheme is formidably expensive. Here, we propose a formalism based on the kernel polynomial method (KPM), which makes the calculation dramatically efficient. We perform a benchmark calculation for bcc-Fe, fcc-Co, and fcc-Ni and find that the effect of the magnetic non-local terms is particularly prominent for bcc-Fe. We also present several local approximations to the magnetic non-local terms for which we can apply the Green's function method and reduce the numerical cost further by exploiting the intermediate representation of the Green's function. By comparing the results of the KPM and local methods, we discuss which local method works most successfully. Our approach provides an efficient way to estimate the Curie temperature of metallic magnets with a complex spin configuration.
\end{abstract}

\maketitle

\section{Introduction}
Non-empirical calculation of the transition temperature ($T_c$) of magnets is one of the long-standing challenges in condensed-matter physics. In particular, it has been well known that the problem becomes extremely difficult and highly non-trivial when the system is metallic. To cope with this problem, there are two possible approaches. One is based on the {\it ab initio} downfolding method, in which we first derive an effective Hamiltonian for the itinerant low-energy electrons~\cite{Gunnarsson1976, Reser1999},  and then accurately solve the model by a sophisticated many-body method such as the dynamical mean-field theory (DMFT)~\cite{Lichtenstein2001, Belozerov2013, Belozerov2017, Mravlje2012, Poteryaev2016, Okamoto2017}. 
However, due to its expensive numerical cost, it is still a formidable task to calculate $T_c$ of magnets with a complex magnetic structure.

The second approach starts with the mapping to an effective spin model in which we focus on the spin degrees of freedom of the system~\cite{Wang1982, Oguchi1983, Lichtenstein1984, Lichtenstein1985, Gyorffy1985, Staunton1986, Liechtenstein1987, Sandratskii1989a, Sandratskii1989b, Staunton1992, Uhl1996, Mryasov1996, Halilov1998, Bruno2003}.
Here, the so-called local force method has been widely used. This method is based on the idea that the energy responses against the spin rotations provide complete information about the exchange interactions in the spin model. The method is applicable regardless of whether the system is metallic or insulating. By combining the spin density functional theory (SDFT), we can derive a spin model without introducing any empirical parameter.

The local force approach was first formulated in the multiple scattering theory with the Green's functions techniques. Thus, it was implemented in SDFT calculations with the Korringa-Kohn-Rostoker (KKR) theory~\cite{Oguchi1983,Lichtenstein1984}. Early studies based on the linear muffin-tin orbital (LMTO) basis~\cite{Anderson1975, Gunnarsson1983, Sabiryanov1995, Sakuma1999} exploited their analogous forms to the KKR equations. There, the single-site scattering operator and scattering path operator in KKR were replaced by the inverse of the potential function and Green's function, respectively. This technique has been successful in estimating $T_c$ of a variety of systems, including non-collinear magnets~\cite{Kubler1988, Sakuma2000} and magnetic alloys~\cite{Takahashi2007}.
%with the help of the coherent potential approximation~\cite{Takahashi2007}.
%However, since the formulation highly relies on the properties of its specific basis, it is not easy to replace them by the other modern DFT basis, such as the linearized augmented plane wave. 

Recently, the local force method has been applied to the SDFT Hamiltonian with various spatially localized bases such as the LMTO~\cite{Katsnelson2000, Kvashnin2015, Kvashnin2016}, linear-combination of pseudo atomic orbital (LCPAO)~\cite{Yoon2018, Terasawa2019}, and Wannier orbital~\cite{Korotin2015}. Especially, the Wannier-based approach has the broadest applicability, since one can construct Wannier functions irrespective of the choice of the basis of the SDFT calculation~\cite{Marzari1997, Marzari2012, w90}. This is a great advantage when we perform a large-scale calculation for magnets with many magnetic atoms in the unit cell using the plane-wave basis.

However, there is a serious drawback of the Wannier-based approach: The derived tight-binding model always contains spin-dependent transfer terms, i.e., non-local magnetic potential terms. While such non-local terms are crucial for the self-consistent mapping to the effective spin model, the numerical cost to take account of them is extraordinarily expensive. Thus the effect of these terms has yet to be investigated in the previous studies~\cite{Korotin2015, Terasawa2019}. In this paper, we present a formalism using the kernel polynomial method (KPM), which is known as a real-space solver for the bilinear Hamiltonian~\cite{Silver1994, Weibe2006}. We show that the numerical cost is dramatically reduced, and the calculation including magnetic non-local terms becomes feasible.

We then apply the present method to bcc-Fe, fcc-Co, and fcc-Ni. We find that the effect of magnetic non-local terms on $T_c$ is prominent for bcc-Fe. We also present several local approximations to the magnetic non-local terms in the Green's function formalism. There, with the help of the intermediate representation of the Green's function~\cite{Shinaoka2017,Chikano2019}, the calculation becomes more efficient, especially at low temperatures. By comparing the results of the KPM and local approximation methods, we also discuss which local approximation successfully reproduces the KPM result. These results will pave an efficient way to evaluate $T_c$ of metallic magnets with a complex magnetic structure such as a skyrmion crystal~\cite{Nagaosa2013}.

\section{Formulation}
In this section, we summarize the formulation of the local force method for the tight-binding model using the Wannier basis. %Although part of them are found in the previous works such as Ref. \cite{Katsnelson2000}, we repeat them here for the sake of self-containedness. 

\subsection{Tight-binding model}
We start with the following tight-binding Hamiltonian $\cH$ in the Wannier representation,
\begin{align}
\cH=\sum_{12}A_{12} c_1^\dagger c_2, \label{eq:first}
\end{align}
where the indices $1,2$ run over all degrees of freedom that specify the Wannier functions, namely, lattice vectors, sublattices, atomic or molecular orbitals, and spins~\cite{memo01}. $A_{12}$ denotes a hopping integral matrix and $c^\dagger_1$ ($c_1$) is an electron creation (annihilation) operator in this basis. 

DFT Hamiltonian leading a Kohn-sham equation takes a form of $\cH=\int d\br\; \vec{\psi}^\dagger(\br)h(\br,{\bm \partial})\vec{\psi}(\br)$ where $h(\br,{\bm \partial})$ is a single-particle Hamiltonian matrix and $\vec{\psi}^\dagger(\br)$ ($\vec{\psi}(\br)$) is a spinor field creation (annihilation) operator. By expanding $\vec{\psi}(\br)$ by a set of Wannier functions $\{\vec{w}_1(\br)\}$, we see that $A_{12}$ is given by,
\begin{align}
A_{12}=\int d\br\; \vec{w}_{1}^\dagger(\br)h(\br,{\bm \partial})\vec{w}_{2}(\br). \label{eq:A12}
\end{align}
In LSDA, $h(\br,{\bm \partial})$ generally consists of the non-magnetic and magnetic parts as follows:
\begin{align}
h(\br,{\bm \partial})=h_0(\br,{\bm \partial})+g\mu_B{\bm B}_{\rm eff}(\br)\cdot {\bm \sigma},
\end{align}
where the second term breaks time-reversal symmetry while the first term preserves it. Here, ${\bm B}_{\rm eff}(\br)$ represents a effective magnetic field due to the magnetic order and is parallel to the ordered moment~\cite{Kleinman1999, Capelle2001}. One may separate $A_{12}$ into $t_{12}$ and $v_{12}$ according to the time-reversal symmetry, and then, these would become, 
\begin{align}
t_{12}&=\int d\br\; \vec{w}_{1}^\dagger(\br)h_0(\br,{\bm \partial})\vec{w}_{2}(\br), \label{eq:t12}\\
v_{12}&=g\mu_B\int d\br\;  \vec{w}_{1}^\dagger(\br)({\bm B}_{\rm eff}(\br)\cdot{\bm \sigma}) \vec{w}_{2}(\br).\label{eq:v12}
\end{align}
We call $v_{12}$ in Eq.~\eqref{eq:v12} the magnetic potential term hereafter. It should be noted that the magnetic order generally deforms the shape of the Wannier functions differently for up and down spins. However, here we assume that this effect is negligibly small, and the time-reversal symmetry breaking term is given by Eq.~\eqref{eq:v12}. In the following, we only consider the cases without spin-orbit coupling, and $t_{12}$ becomes the identity matrix in the spin space.
%, in the conventional downfolding process, \hr{resulting $t_{12}$ and $v_{12}$ are not exactly identical} to Eqs.~\eqref{eq:t12} and \eqref{eq:v12}, respectively, since the magnetic order generally deforms the shape of the Wannier functions and $t_{12}$ in Eq.~\eqref{eq:t12} no longer respects the time-reversal symmetry. 

\subsection{Spin model}
In the local force approach, we map the original itinerant models to the classical spin models defined as follows:
\begin{align}
\cH_{\rm SM}=-2\sum_{\braket{i,j}}J_{ij}\be_i\cdot\be_j, 
\end{align}
where $i,j$ specify atomic sites (namely, lattice vectors and sublattices), $\be_i$ is a local spin moment normalized to $|\be_i|=1$, and $J_{ij}$ is the exchange interaction between two spins. The summation runs over the interacting bonds, where the self-interaction terms, $J_{ii}$, are excluded. Here, we choose the simplest Heisenberg model as the mapped spin system, which includes only bilinear terms of the exchange interactions. The higher-order exchange interactions, Dzyaloshinskii-Moriya interaction, and magnetic anisotropy in the presence of spin-orbit coupling can be taken into account by slight modifications~\cite{Mryasov1996, Katsnelson2000}. 

Following Refs.~\cite{Oguchi1983,Lichtenstein1984}, let us consider the excitation energies by rotating the magnetic moments in the collinear ferromagnets, where all spins are along the $z$-direction. For the spin rotation at $i$-site of the angle $\theta_i$, the relation $\delta^2 E_{\rm SM}/\delta \theta^2_i=2\sum_{j\neq i}J_{ij}$ holds at every $i$-site. On the other hand, the two spin rotation at $i$-site of the angle $\theta_i$ and  $j$-site of the angle $\theta_j$ leads to the identity $\delta^2 E_{\rm SM}/\delta \theta_i\delta \theta_j=-2J_{ij}$. Thus, the following relation, 
\begin{align}
\frac{\delta^2 E_{\rm SM}}{\delta \theta_i^2}=-\sum_{j\neq i} \frac{\delta^2 E_{\rm SM}}{\delta \theta_i\delta\theta_j},\label{eq:sum}
\end{align}
holds for the collinear ferromagnet in the classical Heisenberg model. This is a kind of sum rule that should be satisfied not only in the mapped spin model but also in the original itinerant system in the local force approach. 

If the unit cell contains only one magnetic atom, $J_0=\sum_{j\neq i}J_{ij}$ does not depend on the site $i$, and then, the mean field value of $T_c$ is given by,
\begin{align}
T_c=\frac{2}{3}J_0. \label{eq:mftc}
\end{align}
While Eq.~\eqref{eq:mftc} often overestimates $T_c$ in real materials, let us focus on $J_0$ hereafter. 

\subsection{Spin rotation in the tight-binding model}
Here, we consider the effect of spin rotation in the itinerant tight-binding model to map it to the spin system. Unfortunately, the definition of spin rotation itself is not obvious in the itinerant model since the localized spin picture no longer holds. In the KKR formalism, spin rotation is expressed by the rotation of the single-site scattering matrix $t(\varepsilon)$, and then, the sum rule \eqref{eq:sum} is automatically satisfied~\cite{Liechtenstein1987}. Calculation with LMTO can exploit its formal similarity with KKR and respects the sum rule: The LMTO eigenvalue equation becomes an equivalent form to that in KKR by neglecting the non-orthogonality of the LMTO basis, in which $t^{-1}(\varepsilon)$ in KKR is replaced by the potential function 
%$P(\varepsilon)$ 
in LMTO. 

On the other hand, in the formulation based on the tight-binding model, one usually regards the spin rotation as the rotation of magnetic potential terms in the Hamiltonian ($v_{12}$ in this paper)~\cite{Katsnelson2000, Kvashnin2015, Kvashnin2016, Korotin2015, Yoon2018, Terasawa2019}. If they are local quantities and do not have site off-diagonal components, the sum rule~\eqref{eq:sum} will be satisfied~\cite{Katsnelson2000}. However, in the tight-binding model constructed from SDFT, there is no justification that the site off-diagonal components of $v_{12}$ are negligibly small. Indeed, it is necessary to consider them to reproduce the original band structure of SDFT. Note that such a difficulty does not appear when we perform the DMFT calculation with the on-site Hubbard interactions since the resulting magnetic potential becomes a local quantity~\cite{Katsnelson2000}. However, we would face the same problem once we consider a momentum-dependent self-energy to improve DMFT.  

Here, we show that the above difficulty due to the site off-diagonal elements of $v_{12}$ is formally eliminated by decomposing $v_{12}$ into the contribution of each site $i$: 
\begin{align}
v_{12}=\sum_i v_{12}^i.\label{eq:sepv}
\end{align}
Then, we can express the $i$-site spin rotation as the rotation of $v_{12}^i$ as follows: Let $v_{12}(\bn_i, \theta_i)$ denotes the magnetic potential, where the $i$-site spin is rotated along ${\bm n}_i$-axis by the angle $\theta_i$ from the original magnetic structure. One may define $v_{12}(\bn_i, \theta_i)$ by the following equation:
\begin{align}
v_{12}(\bn_i, \theta_i)=[D^\dagger(\bn_i,\theta_i)v^i D(\bn_i,\theta_i)]_{12}+\sum_{j\neq i}v_{12}^j,
\end{align}
where $D_{12}(\bn_i,\theta_i)$ is expressed by the rotation matrix for $s=1/2$ spinor basis,  $D_{\sigma_1\sigma_2}(\bn_i,\theta_i)=[e^{-i\frac{\theta_i}{2}(\bn_i\cdot{\bm \sigma})}]_{\sigma_1\sigma_2}$, as $D_{12}(\bn_i,\theta_i)=\delta_{i_1i_2}\delta_{\ell_1,\ell_2}D_{\sigma_1\sigma_2}(\bn_i,\theta_i)$. Here, we have used the symbolic notation, $1=(i \ell \sigma)_1=(i_1,\ell_1,\sigma_1)$, where $\ell$ and $\sigma$ respectively represent the orbital and spin degrees of freedom of the Wannier function~\cite{memo02}.

With the above setup, the deviation $\delta v_{12}(\bn_i, \theta_i)=v_{12}(\bn_i, \theta_i)-v_{12}$ can be expanded by $\theta_i$ as follows: 
\begin{align}
\delta v_{12}(\bn_i, \theta_i)=\Sigma^{i(0)}_{12}\theta_i+\Sigma^{i(1)}_{12}\theta_i^2+\cO(\theta_i^3),
\end{align}
where $\Sigma^{i(0)}$ and $\Sigma^{i(1)}$ are defined by,
\begin{align}
\Sigma^{i(0)}&=-\frac{i}{2}\bn_i\cdot [v^i,{\bm \sigma}]_-,\\
\Sigma^{i(1)}&=\frac{1}{4}((\bn_i\cdot{\bm \sigma})v^i(\bn_i\cdot{\bm \sigma})-v^i).
\end{align}
Here, $[A,B]_\pm=AB\pm BA$. The Pauli's matrix ${\bm \sigma}$ only acts on the spin index, namely, ${\bm \sigma}_{12}=\delta_{i_1i_2}\delta_{\ell_1\ell_2}{\bm \sigma}_{\sigma_1\sigma_2}$.

Here, we consider the possible forms of $v_{12}^i$. Since the definition \eqref{eq:sepv} has large ambiguity, we have to choose an appropriate form depending on the situation. For example, if the site off-diagonal elements of $v_{12}$ are exactly zero, we can simply set,
\begin{align}
v_{12}^i=\delta_{i i_1}\delta_{i_1i_2}v_{12}.
\end{align}
In the case that $\ell_1$ orbital is much more localized than $\ell_2$ orbital, like a $c$-$f$ hybridization in rare-earth compounds, the dominant contribution of Eq.~\eqref{eq:v12} comes form the region close to $i_1$-site. Thus, the following type separation would be physically reasonable:
\begin{align}
v_{12}^i = \left\{\begin{aligned}
\delta_{ii_1}v_{12}\quad\mbox{if $\ell_1$ is more localized}\\
\delta_{ii_2}v_{12}\quad\mbox{if $\ell_2$ is more localized}\\
\end{aligned}\right..
\end{align}
On the other hand, in the case that both $\ell_1$ and $\ell_2$ orbitals equally contribute Eq.~\eqref{eq:v12}, we may choose,
\begin{align}
v_{12}^i = \frac{1}{2}(\delta_{i i_1}+\delta_{ii_2})v_{12}, \label{eq:v12sep}
\end{align}
as the separation. In this paper, we simply use Eq.~\eqref{eq:v12sep} and leave how other choices affect the estimation of $T_c$ for a future work.

\subsection{Green's function formalism}
In the conventional Green's function formalism, the free energy $F$ of the system \eqref{eq:first} is evaluated by,
\begin{align}
F= -T\sum_{\omega_n}e^{i\omega_n 0^+}{\rm Tr}\ln[-\beta G^{-1}(i\omega_n)].
\end{align}
Here, the trace ${\rm Tr}$ runs over all indices, and we have introduced an infinitesimal positive constant $0^+$ to guarantee the convergence~\cite{note03}. The Green's function $G(i\omega_n)$ is defined by $G^{-1}_{12}(i\omega_n)=(i\omega_n\delta_{12}-A_{12})$. By using standard perturbation techniques, we can evaluate $\delta^2 F/\delta\theta_i^2$ for the one spin rotation and $\delta^2 F/\delta\theta_i\delta \theta_j$ for the two spin rotation as follows:
\begin{align}
\frac{\delta^2 F}{\delta\theta_i^2}&=2T\sum_{\omega_n}e^{i\omega_n 0^+}{\rm Tr}[G(i\omega_n)\Sigma^{i(1)}]\nonumber\\&\hspace{1cm}+T\sum_{\omega_n}{\rm Tr}[G(i\omega_n)\Sigma^{i(0)}G(i\omega_n)\Sigma^{i(0)}],\label{eq:lich1}\\
\frac{\delta^2 F}{\delta\theta_i\delta\theta_j}&=T\sum_{\omega_n}{\rm Tr}[G(i\omega_n)\Sigma^{i(0)}G(i\omega_n)\Sigma^{j(0)}].\label{eq:lich2}
\end{align}
Equations \eqref{eq:lich1} and \eqref{eq:lich2}, or their analytic continuations, are the so-called Lichtenstein's formula to evaluate $J_{ij}$ by using the Green's functions. 

Let us consider again the collinear ferromagnetic order with $z$-axis polarization. In this case, we can set ${\bm n}_i$ to the $y$-axis, and then, $\Sigma^{i(0)}$ and $\Sigma^{i(1)}$ become,
\begin{align}
\Sigma^{i(0)}=\tilde{v}^i\sigma^x,\;\;\mbox{and}\;\;
\Sigma^{i(1)}=-\frac{1}{2} \tilde{v}^i\sigma^z,
\end{align}
where we define $\tilde{v}^i$ as $v_{(i\ell \sigma)_1(i\ell \sigma)_2}^i=\tilde{v}_{(i\ell)_1(i\ell)_2}^i\sigma^z_{\sigma_1\sigma_2}$. Similarly, the Green's function $G(i\omega_n)$ becomes diagonal in the spin space, whose $\sigma$-$\sigma$ submatrix $G^\sigma(i\omega_n)$ is given by $[G^{\sigma}(i\omega_n)]^{-1}=i\omega_n\delta - \tilde{t}-\sigma\tilde{v}$, where $\sigma=1$ ($-1$) denotes the spin up (down) component. By using these relations, we can prove the following relation: 
\begin{align}
{\rm Tr}[G\Sigma^{i(1)}]&=-\frac{1}{2}{\rm Tr}_{i\ell}[(G^\uparrow - G^\downarrow)\tilde{v}^i]\nonumber\\
&=-\sum_{j}{\rm Tr}_{i\ell}[G^\downarrow \tilde{v}^j G^\uparrow \tilde{v}^i], \label{eq:dyson}
\end{align}
by inserting the identity $G^\sigma (G^\sigma)^{-1}=1$ and using Eq.~\eqref{eq:sepv}. Here, ${\rm Tr}_{i\ell}$ denotes the trace for $i$ and $\ell$ indices. On the other hand, one can confirm that the relation ${\rm Tr}[G\Sigma^{i(0)}G\Sigma^{i(0)}]=2{\rm Tr}_{i\ell}[G^\downarrow \tilde{v}^i G^\uparrow \tilde{v}^i]$ also holds in the collinear cases, and thus, the following sum rule is satisfied in this approach:
\begin{align}
\frac{\delta^2 F}{\delta \theta_i^2}=-\sum_{j\neq i} \frac{\delta^2 F}{\delta \theta_i\delta\theta_j}.\label{eq:lichsum}
\end{align}
Equation~\eqref{eq:lichsum} is what we desire to guarantee the self-consistency of the mapping. We emphasize here that the decomposition \eqref{eq:sepv} is essential to prove it. 

Unfortunately, Eqs.~\eqref{eq:lich1} and \eqref{eq:lich2} are not so efficient forms in practical calculations when the site off-diagonal component of $v_{12}^i$ is finite. If we can chose local $v_{12}^i$, for example, Eq.~\eqref{eq:lich2} becomes,
\begin{align}
\frac{\delta^2 F}{\delta \theta_i\delta\theta_j}=T\sum_{\omega_n}{\rm Tr}_{\ell \sigma}[G_{ji}(i\omega_n)\Sigma^{i(0)}G_{ij}(i\omega_n)\Sigma^{j(0)}].\label{eq:locallich}
\end{align}
Here, ${\rm Tr}_{\ell\sigma}$ runs over only orbital and spin spaces, and thus, Eq.~\eqref{eq:locallich} can be evaluated with $\cO(N_{\ell\sigma}^3 N_M)$ operations, where $N_{\ell\sigma}$ is the dimension of the orbital and spin space, and $N_M$ is the maximum number of the Matsubara frequency. However, when the off-diagonal $v_{12}^i$ remains finite, we have to take a trace not only for the orbital and spin space but also for the site space, which makes a evaluation of Eq.~\eqref{eq:lich2} prohibitively difficult in complex multi-orbital systems. Since the spin rotation, thus $\Sigma^{i(0)}$, breaks the lattice translation symmetry, Fourier transformation to the momentum space does not reduce the computational cost. After all, it requires $\cO(N^3N_M)$ operations where $N$ is the dimension of the hopping integral matrix $A_{12}$. To make the calculation feasible, here, we propose the following two ways:
\begin{enumerate}
\item[(i)] To approximate $v_{12}$ as a local quantity, and evaluate Eqs.~\eqref{eq:lich1} and \eqref{eq:lich2}. In this paper, we perform three calculations along this line, two of which violate the sum rule~\eqref{eq:lichsum} while the rest does not (see, (A), (B), and (C) approaches in Sec.~IV~B). %It should be noted that we can neglect the off-site components only of $\Sigma^{i0}/\Sigma^{i1}$ but consider those of $v_{12}$ in $G_{12}(i\omega_n)$, in which the sum rule \eqref{eq:lichsum} no longer holds. However, as is shown later, this procedure of computing $T_c$ works better than that neglecting the off-site components of $v_{12}$ in the Hamiltonian level for which Eq.~\eqref{eq:lichsum} is exactly satisfied (see, (A), (B), and (C) approaches in Sec.~IV~B).
\item[(ii)] To evaluate $\delta^2 F/\delta \theta^2_i$ and $\delta^2 F/\delta \theta_i\delta \theta_j$ by other diagonalization technique.
%as is often performed in real space calculations, instead of using the Green's functions. 
Below, we develop a KPM-based scheme which is suitable for calculation in the real space. Although this method still requires a much higher cost than (a), it can estimate $J_0$ without introducing local approximation for $v_{12}$.
\end{enumerate}

\subsection{Kernel polynomial method}
In this subsection, we present a formulation of the local force approach based on KPM. KPM is a kind of sparse matrix diagonalization technique such as the Lanczos algorithm and has often been used to calculate physical quantities in the systems without the translation symmetry~\cite{Silver1994, Weibe2006}. Recently, Barros and Kato applied KPM to the Langevin simulations of the classical Kondo lattice model~\cite{Barros2013}. They studied the chiral domain formation in the triangular lattice system, which was achieved with an efficient computational scheme to evaluate the first derivatives of the free energy $F$. More recently, many techniques have been proposed to improve their method and applied to investigate exotic phenomena such as the formation of the skyrmion crystal~\cite{Barros2014, Ozawa2017, Wang2018}. 

Here, we show that not only the first derivatives but also the second derivatives of $F$ can be evaluated by KPM within the same computational cost as $F$ itself.  Thus, one can apply this technique to the estimation of $T_c$ in the local force approach by evaluating $\delta^2 F/\delta \theta^2_i$ and $\delta^2 F/\delta \theta_i\delta \theta_j$. Here, we start with the following form of the free energy $F$ in KPM:
\begin{align}
F=\sum_{m=0}^{M-1}c_m r^\dagger \alpha_m, \label{eq:kpm0}
\end{align}
which is evaluated as an ensemble average over a random column vector $r$ with order $N$. The coefficients $c_m$ are expressed by the Chebyshev polynomials $T_m(x)=\cos(m \,{\rm arccos}\, x)$, the kernel damping factor $g_m$, and $f(x)=-T\log (1+e^{-\beta(x-\mu)})$ as follows:
\begin{align}
c_m = \frac{1}{\pi}(2-\delta_{0m})g_m\int_{-1}^{1}dx \frac{T_m(x)f(x)}{\sqrt{1-x^2}}.
\end{align}
Here, we choose the Jackson kernel as $g_m$:
\begin{align}
g_m&=\frac{(M-m+1)\cos\frac{\pi m}{M+1}+\sin\frac{\pi m}{M+1}\cot\frac{\pi}{M+1}}{M+1}.
\end{align}
On the other hand, the column vector $\alpha_m$ is defined by the following recursive relations:
\begin{align}
\alpha_m = \left\{\begin{aligned}
r\quad(m=0)\\
Ar\quad(m=1)\\
2A\alpha_{m-1}-\alpha_{m-2}\quad (m\ge2)
\end{aligned}\right..\label{eq:alpha_}
\end{align}
Here, $A$ is the $N\times N$ hopping integral matrix defined in Eq.~\eqref{eq:first}.
It should be noted that $A$ has only $\cO(N)$ finite elements because there are not so many distant hoppings in the tight-binding model.
$\alpha_m$ is obtained from $m=0$ to $M-1$ by using Eq.~\eqref{eq:alpha_}. Since Eq.~\eqref{eq:alpha_} only involves a matrix-vector product, this recursive procedure requires only $\cO(MN)$ operations.

\subsubsection{First derivatives of the free energy}
A remarkable aspect of KPM is that one can calculate the first derivatives of $F$ by the similar recursive procedure as $F$, as is shown in Ref.~\cite{Barros2013}. According to their results, the first derivatives of $F$ by $A_{12}$ are given as follows:
\begin{align}
\frac{\partial F}{\partial A_{12}}=2\sum_{m=0}^{M-2}[\beta_m]_1 [\alpha_m]_2.\label{eq:kpm1}
\end{align}
Here, the column vector $\beta_m$ is calculated from $\beta_m=0$ for $m\ge M-1$, followed by,
\begin{align}
\beta_m = c_{m+1}r^\dagger+2\beta_{m+1}A-\beta_{m+2}, \label{eq:kpmbeta}
\end{align}
for $M-2 \ge m \ge 1$, and $\beta_0 = \frac{1}{2}(c_{1}r^\dagger+2\beta_{1}A-\beta_{2})$. Since $\beta_m$ and $\alpha_m$ require only the matrix-vector products in Eqs.~\eqref{eq:alpha_} and \eqref{eq:kpmbeta}, we can simultaneously obtain all components of $\delta F/\delta A_{12}$ with the $\cO(MN)$ operations. A summary of the derivation is given in the appendix. 

\subsubsection{Second derivatives of the free energy}
Similar to the first derivatives, we can derive the formulas for the second derivatives of $F$ with the recursive relations. The details of the derivation are found in the appendix and the results become,
\begin{align}
\hspace{-1mm}\frac{\partial^2 F}{\partial A_{12} \partial A_{34}}=4\sum_{m=1}^{M-2}[\beta_m]_3 [\gamma_m^{12}]_4+4\sum_{m=0}^{M-3}[\tilde{\gamma}_m^{12}]_3 [\alpha_m]_4, \label{eq:kpm2}
\end{align}
where the row vector $\gamma^{12}_m$ and the column vector $\tilde{\gamma}^{12}_m$ respectively satisfy the following recursive forms, 
\begin{align}
\gamma_m^{12}&=\Delta^{12}\alpha_{m-1}+2A\gamma^{12}_{m-1}-\gamma^{12}_{m-2}, \label{eq:gamma1}\\
\tilde{\gamma}_m^{12}&=\beta_{m+1}\Delta^{12}+2\tilde{\gamma}_{m+1}^{12}A-\tilde{\gamma}_{m+2}^{12}. \label{eq:gamma2}
\end{align}
Here, the matrix $\Delta^{12}$ is defined by $[\Delta^{12}]_{34}=\delta_{13}\delta_{24}$. Using Eq.~\eqref{eq:gamma1} (Eq.~\eqref{eq:gamma2}), we can calculate $\gamma_m^{12}$ ($\tilde{\gamma}_m^{12}$) starting with $\gamma_m^{12}=0$ for $m\le 0$ ($\tilde{\gamma}_m^{12}=0$ for $m\ge M-2$). Then, the desired second derivatives in KPM are obtained by the following chain rule,
\begin{align}
\frac{\delta^2 F}{\delta \theta_i\delta \theta_j}&=\sum_{12}\frac{\delta^2 A_{12}}{\delta \theta_1\delta \theta_2}\frac{\delta F}{\delta A_{12}}+\sum_{1234}\frac{\delta A_{12}}{\delta \theta_1}\frac{\delta A_{34}}{\delta \theta_2}\frac{\delta^2 F}{\delta A_{12}\delta{A_{34}}} \label{eq:kpm2_}
\end{align}
Combining Eqs.~\eqref{eq:kpm1}-\eqref{eq:kpm2_}, we finally obtain the following formulas:
\begin{align}
\frac{\delta^2 F}{\delta \theta_i^2}&=4\sum_{m=0}^{M-1}{\rm Tr}\left[\Sigma^{i(1)} (\beta_m\otimes\alpha_m)^T\right]\nonumber\\
&+4\left[ \sum_{m=1}^{M-2} \braket{\beta_m\Sigma^{i(0)}\Gamma_m^i}+\sum_{m=0}^{M-3} \braket{\tilde{\Gamma}_m^i\Sigma^{i(0)}\alpha_m}\right], \label{eq:kpm_res1}\\
\frac{\delta^2 F}{\delta \theta_i \delta \theta_j}&=\hspace{-0.5mm}4\left[ \sum_{m=1}^{M-2} \braket{\beta_m\Sigma^{j(0)}\Gamma_m^i}+\sum_{m=0}^{M-3} \braket{\tilde{\Gamma}_m^i\Sigma^{j(0)}\alpha_m}\right].\label{eq:kpm_res2}
\end{align}
Here, we have used symbolic notations $[(\beta_m\otimes \alpha_m)^T]_{12}=[\beta_m]_2[\alpha_m]_1$, $\braket{\beta_m\Sigma^{i(0)}\Gamma^i_m}=\sum_{12}[\beta_m]_1\Sigma^{i(0)}_{12}[\Gamma^i_m]_{2}$, and so on. The row and column vectors $\Gamma_m^i$ and $\tilde{\Gamma}_m^i$ are defined by $\gamma_m$ and $\tilde{\gamma}_m$, respectively, as follows:
\begin{align}
\Gamma_m^i=\sum_{12}\Sigma_{12}^{i(0)}\gamma_m^{12}, \;\;\mbox{and} \;\;\ \tilde{\Gamma}_m^i=\sum_{12}\Sigma_{12}^{i(0)}\tilde{\gamma}_m^{12}.
\end{align}
These are evaluated also in the recursive forms:
\begin{align}
\Gamma^i_m&=\Sigma^{i(0)}\alpha_{m-1} + 2A\Gamma_{m-1}^{i}-\Gamma_{m-2}^i,\label{eq:kpmGamma1}\\
\tilde{\Gamma}_m^i&=\beta_{m+1}\Sigma^{i(0)} + 2\tilde{\Gamma}_{m+1}^iA-\tilde{\Gamma}_{m+2}^i,\label{eq:kpmGamma2}
\end{align}
which are similar to $\gamma_m$ and $\tilde{\gamma}_m$. Since Eqs.~\eqref{eq:kpm_res1}-\eqref{eq:kpmGamma2} involve only matrix-vector and vector dot products, we can evaluate them without increasing computational cost. Equations \eqref{eq:kpm_res1}-\eqref{eq:kpmGamma2} are one of the main results in this paper. 

\section{Details of the numerical calculation}
Here, we make some remarks on the computational cost in the practical calculations. In the Green's function formalism, we have to evaluate Eq.~\eqref{eq:lich1} and \eqref{eq:lich2} with the approximated $\Sigma^{i(0)}$ and $\Sigma^{i(1)}$, which requires $\cO(N_{\ell\sigma}^3 N_M)$ operations. The required steps can be significantly reduced by using the intermediate representation of the Green's function~\cite{Shinaoka2017,Chikano2019}. There, the Green's function is expanded in terms of the IRbasis~\cite{Chikano2019}, which is a compact basis set that accurately represents an imaginary time dependence of the Green's functions. The number of required basis $N_{\rm IR}$ scales proportionally to $\log W\beta$ where $W$ is the maximum frequency of the energy spectrum and $\beta$ the inverse temperature. Thus, the calculation becomes very efficient at  low temperatures. For typical parameters of $W\sim 10$ eV and $T\sim 0.01$ eV, one finds that $N_{\rm IR}\sim10^2$ is enough to get a convergent solution, which is about two orders of magnitude smaller than $N_M\sim W\beta=10^4$. It should be noted that, in the real frequency representation, another efficient algorithm, called finite pole approximation, to compute Eq.~\eqref{eq:lich2} has been recently proposed~\cite{Terasawa2019,Ozaki2007}. The required number of frequency points is about $10^2$ at $T=300$K, which is comparable to $N_{\rm IR}$. 

In the actual evaluation of Eq.~\eqref{eq:lich2}, we use a sparse sampling approach implemented in IRbasis~\cite{Li2019}. First, we evaluate the following function, 
\begin{align}
P(i\omega_n^F)={\rm Tr}[G(i\omega_n^F)\Sigma^{i(0)}G(i\omega_n^F)\Sigma^{j(0)}], 
\end{align}
for the proper fermionic sampling points $i\omega_n^F$, the number of which is essentially the same as $N_{\rm IR}$. Then, we calculate the coefficients $P_\ell$ of the given basis functions $U_\ell^F(i\omega_n^F)$ by using the least square fitting of $P(i\omega_n^F)=\sum_\ell P_\ell U_\ell^F(i\omega_n^F)$. Finally, we obtain $P(\tau=0)$ by using $P(\tau=0)=\sum_\ell P_\ell U^F_\ell(\tau=0)$. For the details of $U^F_\ell$, see Ref.~\cite{Chikano2019}. Since the above transformation from $P(i\omega_n^F)$ to $P(\tau=0)$ takes much less time than the evaluation of $P(i\omega_n^F)$ itself, the computational cost scales $\cO(N_{\ell\sigma}^3N_{\rm IR})$ if we employ the local approximations for $\Sigma^{i(0)}$ and $\Sigma^{i(1)}$. 

The number of operations required in KPM is estimated as $\cO(SMN)$, where $S$ is the number of elements used in the ensemble average over column vectors $r$. It is known that the required $S$ strongly depends on the complexity of the system, the desired accuracy, and also the probing algorithm of the ensemble. Recently, Wang {\it et al}. proposed an efficient method to choose a proper set of $r$, which they call the optimal coloring technique~\cite{Wang2018}. We apply this method to our multi-orbital systems, and find that $S\sim 3000$ is enough to get converged solutions within the error of $\sim 30$ K in the estimation of $T_c$ for bcc-Fe, which will be discussed in the next section. 

Another relevant parameter in KPM is the number of finite elements in the matrix $A_{12}$. Although the hopping integrals show the exponential decay with respect to the distance, and thus, the number of the finite element is proportional to $N$, its factor can be huge in general multi-orbital systems. Here, we introduce a cutoff energy $\epsilon_{\rm cutoff}$ and neglect matrix elements when the absolute value is lower than $\epsilon_{\rm cutoff}$. 
%Although this is an approximation that is unnecessary in the Green's function formalism, 
As will be shown later, we obtain good convergence with respect to $\epsilon_{\rm cutoff}$.

\section{Results and Discussion}
\subsection{Calculation condition}
The calculations in this paper are organized as follows: First, we perform the SDFT calculation with WIEN2k package~\cite{wien2k}, which implements the full-potential all-electron method based on the linearized augmented plane-wave basis. GGA-PBE exchange correlation functional~\cite{gga}, $R_{\rm MT}K_{\rm max}=9$ of the cutoff parameter, and the number of $\bm k$-points $N_k=10^3$ are employed in the self-consistent calculations. Here, we set the lattice constants as the experimental values $a_{\rm Fe}=2.867$\AA,  $a_{\rm Co}=3.544$\AA, and $a_{\rm Ni}=3.540$\AA. 

The wannierization process is conducted by using {\sc wannier90} package~\cite{w90,w90v3} through wien2wannier interface~\cite{w2w}. The outer and inner windows are set $[-10,40]$ eV and $[-10,5]$ eV with respect to the Fermi energy, respectively. Here, we construct the nine orbital model, which contains one $4s$, five $3d$, and three $4p$ atomic orbitals. Here, we do not minimize the size of Wannier functions but keep the symmetry of the  projection functions. A typical spread of $3d$ Wannier orbitals is about $0.4$\AA$^2$. 

Based on the constructed tight-binding model, we apply the local force method to evaluate $J_0$ in the Green's function method with the local approximations and KPM approach discussed in the previous sections. In the Green's function approach, we use a set of parameters $\Lambda=10^5$ (the cutoff parameter of IRbasis), $N_i=32^3$ (the number of unit cells), and $\beta=200$ eV$^{-1}$. In KPM, we set the parameters as $\epsilon_{\rm cutoff}= 5\times 10^{-3}$ eV, $\beta=50$ eV$^{-1}$, $M=2000$, and $N_i=16^3$ unless these are explicitly mentioned. For the ensemble average, we use $S_0=4^3$ as the number of the colors and gather $S_1=50$ results to obtain the averaged value and the statistical error. Thus, total number of $S$ is $S=S_0S_1=3200$. 

\subsection{Results of Green's function formalism} 
\begin{figure}[t]
\centering
\includegraphics[width=8.5cm]{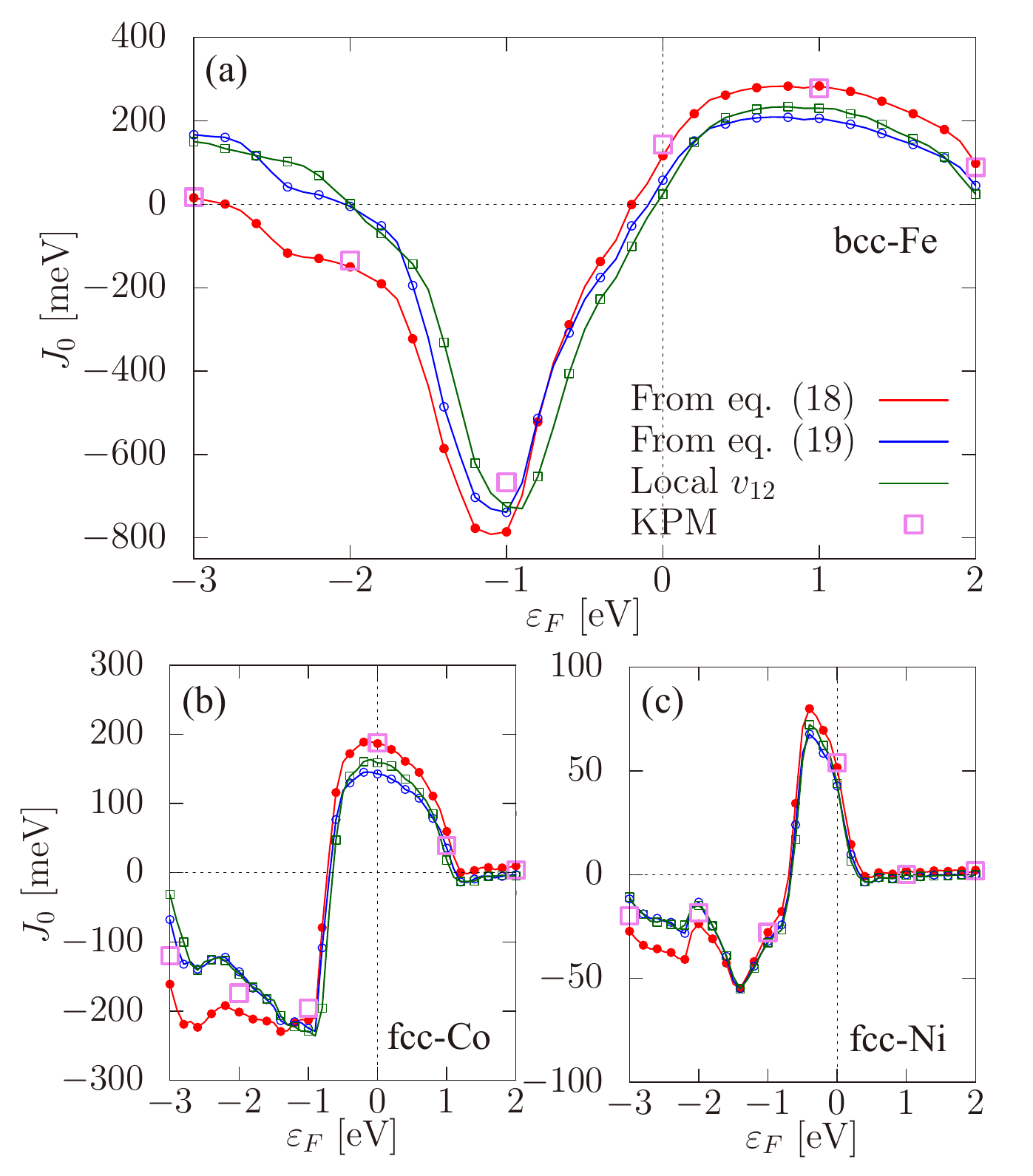}
\caption{Chemical potential dependence of $J_0(\varepsilon_F)$ in (a) bcc-Fe, (b) fcc-Co, and (c) fcc-Ni. Red and blue solid lines indicate the results based on Eqs.~\eqref{eq:lich1} and ~\eqref{eq:lich2} with the approximated $\Sigma^{i(0)}/\Sigma^{i(1)}$, respectively. Green line corresponds to the result where the local approximation for $v_{12}$ is employed. Open violet squares indicate the results of KPM. $\varepsilon_F=0$ corresponds to the actual chemical potential.}
\label{fig:GF} 
\end{figure}

First, we show the results of the Green's function formalism. In Fig.~1(a), we plot $J_0(\varepsilon_F)$ of bcc-Fe as a function of the chemical potential $\varepsilon_F$. 
Here, we shift $\varepsilon_F$ in the tight-binding Hamiltonian and introduce the following three types of approximations:
%to the magnetic non-local terms $v_{12}$. 
\begin{enumerate}
\item[(A)] The red line is the result obtained from $\delta^2 F/\delta \theta_i^2$ with Eq.~\eqref{eq:lich1}. $J_0$ is then calculated using the relation $J_0=(\delta^2 F/\delta \theta_i^2)/2$. 
\item[(B)] The blue line is the result obtained from $\delta^2 F/\delta \theta_i \delta \theta_j$ with Eq.~\eqref{eq:lich2}. $J_0$ is then evaluated using the sum rule \eqref{eq:lichsum}. 
%followed by $J_0=(\delta^2 F/\delta \theta_i^2)/2$.
\end{enumerate}
In these two cases, we employ the local approximation for $v_{12}$ in the calculation of $\Sigma^{i(0)}$ and $\Sigma^{i(1)}$,
neglecting their site off-diagonal components, while full $v_{12}$ is used in the calculation of $G(i\omega_n)$.
\begin{enumerate}
\item[(C)] The green line corresponds to the calculation in which the local approximation for $v_{12}$ is introduced not only to  $\Sigma^{i(0)}$ and $\Sigma^{i(1)}$ but also to $G(i\omega_n)$. Here, the results based on Eqs.~\eqref{eq:lich1} and \eqref{eq:lich2} are identical since the sum rule \eqref{eq:lichsum} is exactly satisfied. 
\end{enumerate}

From Fig.~\ref{fig:GF}(a), we can see that the results of the three calculations behave similarly as a function of $\varepsilon_F$. The overall behavior is also consistent with the previous study with TB-LMTO basis~\cite{Sakuma1999}. 
%However, if we look into the detail, we see that (A) deviates from (B) and (C). 
The fact that (B) agrees well with (C) seems to indicate that $J_0(\varepsilon_F)$ is insensitive to the site off-diagonal components of $v_{12}$ in $G(i\omega_n)$ of Eq.~\eqref{eq:lich2}. This can be understood since the non-local effect of $v_{12}$ in $G(i\omega_n)$ on $J_0(\varepsilon_F)$ is only of the order of $\cO(v_{\rm nn}/W)$, where $v_{\rm nn}$ is a typical energy scale of the nearest-neighbor magnetic potential term. In the case of $3d$-orbitals in bcc-Fe, we find $v_{\rm nn}\sim 0.05$ eV and $W\sim 10$ eV, and thus, $\cO(v_{\rm nn}/W)\sim 5\times 10^{-3}$, which is a negligibly small number.  

However, this does not mean that the site off-diagonal components of $v_{12}$ are indeed irrelevant in the estimation of $J_0(\varepsilon_F)$ since the results of (A), indicated by the red line in Fig.~\ref{fig:GF}(a), shows quantitative difference from (B) and (C). In particular, if we do not shift the chemical potential (i.e., $\varepsilon_F=0$), $J_0^{\rm (A)}=117$ meV is much larger than $J_0^{\rm (B)}=58$ meV and $J_0^{\rm (C)}=25$ meV. 

\begin{figure}[t]
\centering
\includegraphics[width=8.5cm]{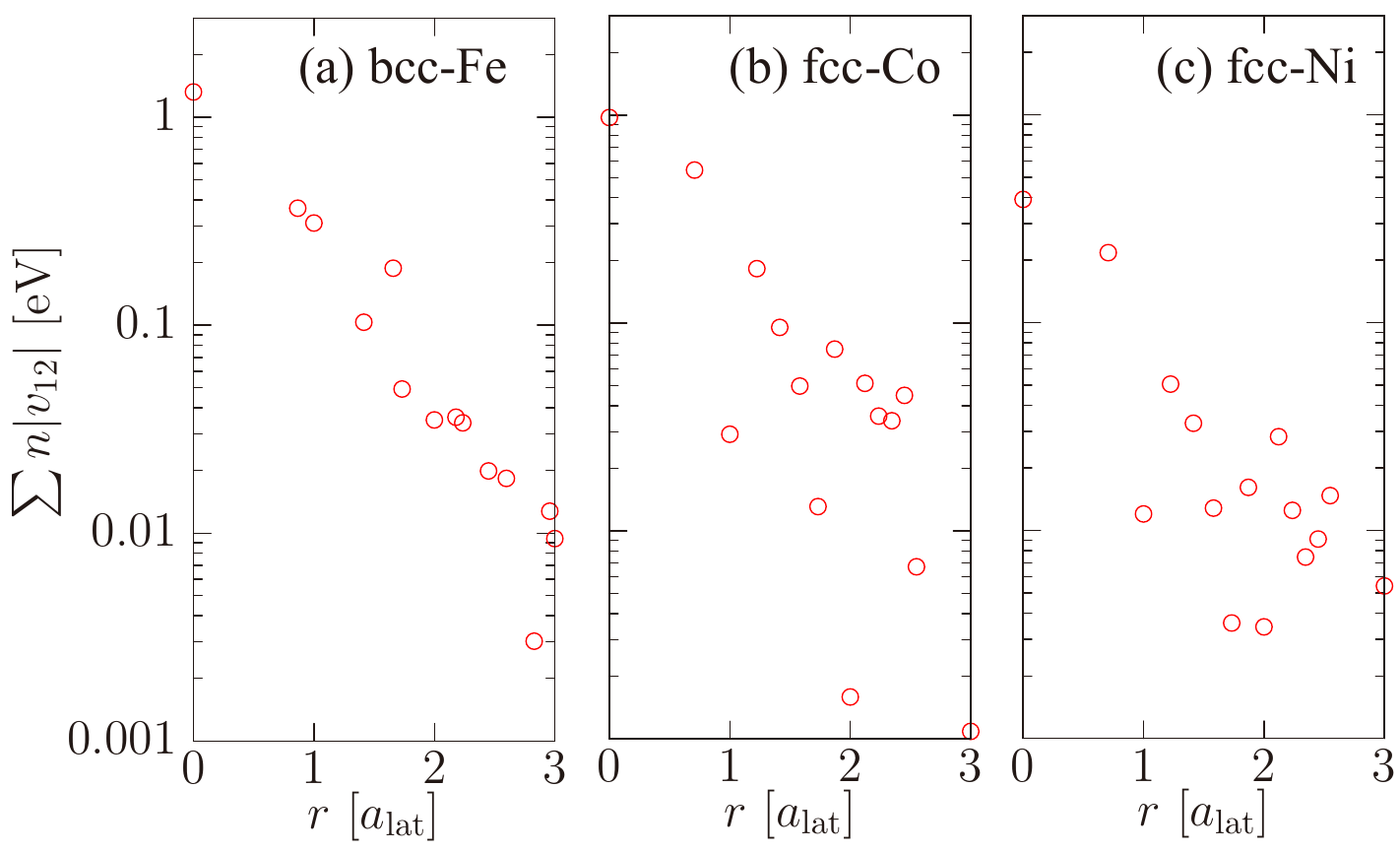}
\caption{Distance dependence of $\sum_{12}n|v_{12}|$ in (a) bcc-Fe, (b) fcc-Co, and (c) fcc-Ni. $n$ denotes the number of equivalent atoms having the same distance.}
\label{fig:hop} 
\end{figure}

To see the origin of this discrepancy more clearly, let us rewrite Eq.~\eqref{eq:lich1} in case (A) as follows:
\begin{align}
\frac{\delta^2 F}{\delta\theta_i^2}&=-2T\sum_{j\neq i} \sum_{\omega_n}{\rm Tr}_{i \ell}[G^\uparrow(i\omega_n)\tilde{v}^{j}G^\downarrow(i\omega_n)\tilde{v}'^{i}]\nonumber \\
&\hspace{5mm}-2T\sum_{\omega_n}{\rm Tr}_{i \ell}[G^\uparrow(i\omega_n)(\tilde{v}^{i}-\tilde{v}'^{i})G^\downarrow(i\omega_n)\tilde{v}'^{i}],\label{eq:lich3}
\end{align}
where $\tilde{v}'^{i}$ denotes the local part of $\tilde{v}^{i}$. Here, we have used the relation \eqref{eq:dyson} and assumed the collinear order. We can see from the first term in Eq.~\eqref{eq:lich3} that (A) partially includes the non-locality of $v_{12}$, which can be estimated from the ratio 
%of the absolute values 
between the local and non-local components of $v_{12}$. In Fig.~\ref{fig:hop}(a), we show the distance dependence of $\sum_{12}|v_{12}|$ of $3d$-orbitals in bcc-Fe, where $r=0$ corresponds to the local one. Since the non-local $v_{12}$ appears as its summation over all sites in the evaluation of Eq.~\eqref{eq:lich3}, here, we multiply $\sum_{12}|v_{12}|$ by $n$, the number of equivalent atoms with the same distance. The result shows that the ratio between the local and non-local components of $v_{12}$ is around $0.3$, which we cannot neglect in the calculations.

The above features are found also in fcc-Co and fcc-Ni, as depicted in Fig.~\ref{fig:GF}(b), (c) and Fig.~\ref{fig:hop}(b), (c). We see that (A) tends to lead larger values of $J_0(\varepsilon_{F}=0)$ than (B) and (C). 

%A similar feature can be found in the previous studies, although we have to keep in mind the difference of the basis functions in the formulation. For example, the recent study by Kvashnin {\it et al.} in Ref.~\cite{Kvashnin2015} shows the mean-field value of $T_c^{\rm LSDA}$ of bcc-Fe as $T_c= 925$ K, which is somewhat smaller than the early results of $T_c=1204$ K in KKR~\cite{Lichtenstein1985} and $T_c=1270$ K in LMTO~\cite{Sabiryanov1995}. This discrepancy may originate from that the calculation in Ref.~\cite{Kvashnin2015} corresponds to the (B) case, while the site off-diagonal effects are effectively included in the latter two calculations.

\subsection{Results of KPM}

\begin{table}
\begin{tabular}{R{1cm} R{1.5cm} R{1.5cm} R{2cm} R{1.5cm}} \hline \hline
& GF$^{\rm (A)}$ & GF$^{\rm (B)}$ & KPM & EXP\\ \hline
Fe& 900 & 448 & 1121$\pm$31 & 1043\\
Co& 1444 & 1104 & 1408$\pm$15& 1388\\
Ni& 399 & 330 & 427$\pm \;\,$6& 627\\ \hline
\end{tabular}
\caption{Mean field value of $T_c$ and the experimental $T_c$ [K]. GF$^{\rm (A)}$ and GF$^{\rm (B)}$ represent the Green's function formalism with the approximations (A) and (B) described in the main text, respectively. EXP denotes the experimental value.}
\label{table}
\end{table}

Next, let us move on to the result of KPM. In contrast with the case of approximation (C) for the Green's function approach, the KPM approach exactly satisfies the sum rule \eqref{eq:lichsum} {\it without} neglecting the non-local magnetic potential terms (i.e., spin-dependent hopping). However, the results of KPM include  statistical error coming from the approximation that replaces the trace of the matrix by the ensemble average over the random vector $r$. Moreover, the approach contains additional parameters $M$ (the number of Chebyshev polynomials in KPM) and $\epsilon_{\rm cutoff}$ (the cutoff energy to the hopping integral matrix $A_{12}$), which control the accuracy of the results and the computational cost. Since the present study is the first application of KPM to the calculation of the second derivatives of the free energy $F$ for the realistic tight-binding model, here we briefly show the $M$ and $\epsilon_{\rm cutoff}$ dependence of $J_0(\varepsilon_{F}=0)$ of bcc-Fe. 
 
\begin{figure}[t]
\centering
\includegraphics[width=8.5cm]{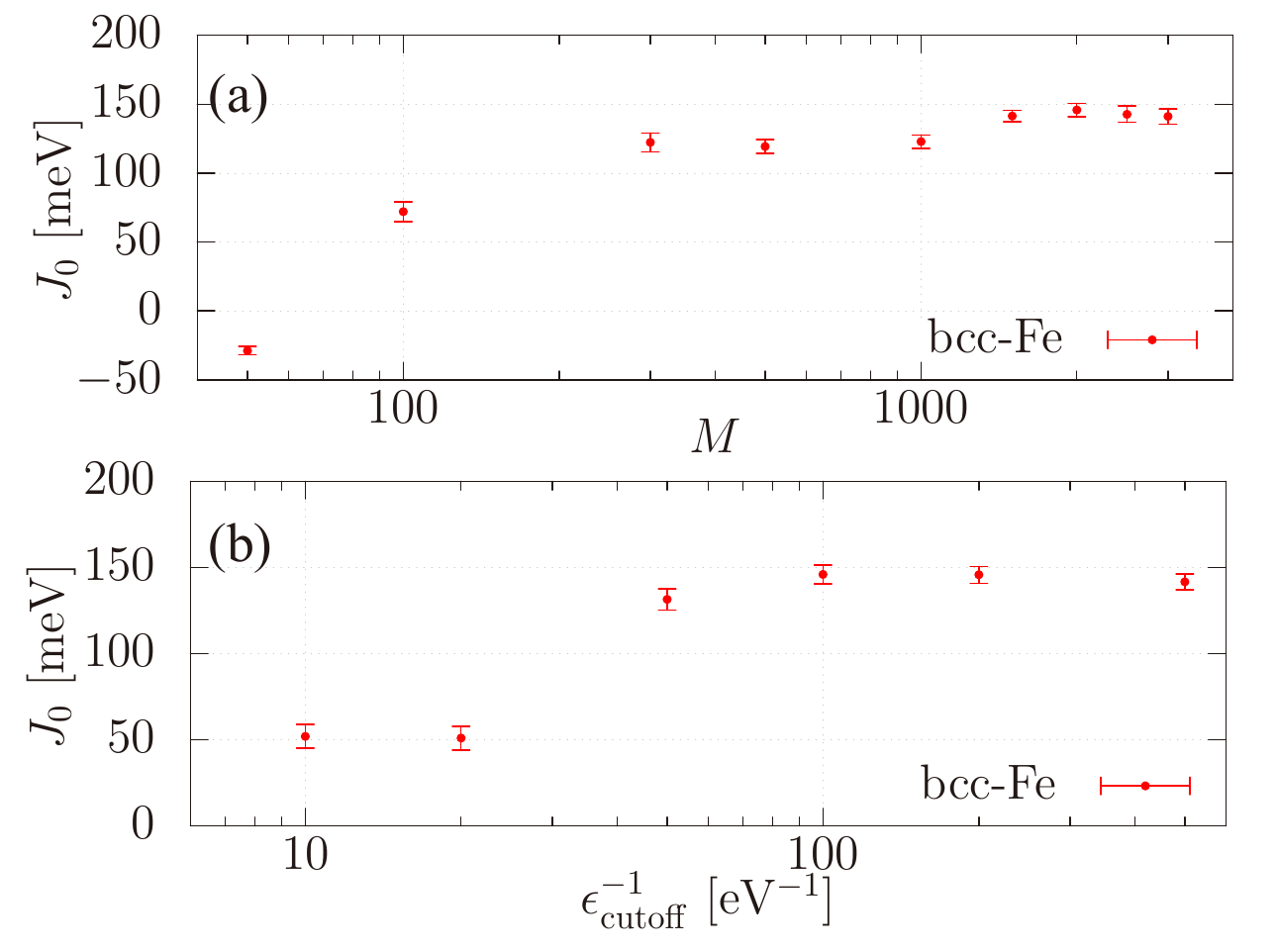}
\caption{Convergence check of $J_0(\varepsilon_{F}=0)$ for bcc-Fe with respect to (a) the number of Chebyshev polynomials $M$, (b) the cutoff energy $\epsilon_{\rm cutoff}$ for the hopping integral $A_{12}$. The other fixed parameters are given in Sec IV A.}
\label{fig:KPM} 
\end{figure}

Figure~\ref{fig:KPM}(a) shows $M$ dependence of $J_0(\varepsilon_{F}=0)$. The required $M$ depends on the temperature and the maximum frequency of the energy spectrum. In our case with $\beta=50$ eV$^{-1}$ and $W\sim10$ eV, we can see that $M=2000$ is enough to obtain the convergent solution within the statistical error. In this sense, the required operation step for the energy direction in KPM is essentially the same as the conventional Matsubara frequency implementation of the Green's function approach.  

Figure~\ref{fig:KPM}(b) shows $\epsilon_{\rm cutoff}$ dependence of $J_0(\varepsilon_{F}=0)$. As can be seen from Fig.~\ref{fig:hop}, the site off-diagonal components of $v_{12}$ are completely ignored when $\epsilon_{\rm cutoff}^{-1}\lesssim 30$ eV$^{-1}$, which will give an unreliable solution. Indeed, from Fig.~\ref{fig:KPM}(b), we can see that the required $\epsilon_{\rm cutoff}^{-1}$ for the convergence is $\epsilon_{\rm cutoff}^{-1}\gtrsim100$ eV$^{-1}$. Note that although the number of finite elements in $A_{12}$ is proportional to $N=N_i\times N_{\ell\sigma}$, its factor strongly depends on $\epsilon_{\rm cutoff}$. At $\epsilon_{\rm cutoff}^{-1}=200$ eV$^{-1}$, for example, it becomes as large as $600$. As a result, the number of the operations with $M=2000$, $S=3200$ and $N_i=16^3$ in KPM is estimated to be $10^{15}$. This is much smaller than $\cO(N^3 N_{\rm M})\sim 10^{18}$ in the conventional Green's function approach with the non-local $\Sigma^{i(0)}$ and $\Sigma^{i(1)}$. It should be noted that we employ $S=3200$ for the sampling point to obtain the statistical error within $30$~K in the estimation of $T_c$ of bcc-Fe. This can be achieved by using the coloring technique in Ref.~\cite{Wang2018}, otherwise the error becomes about $150$ K by using the same number of $S$ with the uniform random vector $r$.
 
The open violet squares in Fig.~\ref{fig:GF} indicate our KPM results, and the corresponding $T_c$ is given in the Table~\ref{table}. The calculated results are consistent with those in the previous studies based on KKR~\cite{Lichtenstein1985} and  LMTO~\cite{Sabiryanov1995}. While the sum rule (Eq.~\eqref{eq:lichsum}) is satisfied and the contribution of spin-dependent hopping in the Wannier representation is effectively considered in these previous studies, let us emphasize here that the present KPM method can always be combined with SDFT calculation regardless of the choice of the basis.

In Fig.~\ref{fig:GF}, we can see that approximation (A) for the Green's function method gives closer values to KPM than (B) and (C), especially when $\varepsilon_F=0$. This implies that, although the sum rule (Eq.~\eqref{eq:lichsum}) is no longer satisfied, (A) works better 
than (B) and (C). This general trend originates from that (A) partially includes the non-local effect of $v_{12}^{i}$, as is discussed in the previous section. 

It should also be noted that
the agreement between KPM and (A) is remarkably good for fcc-Ni and fcc-Co but not so good for bcc-Fe. This result indicates that how the non-local terms affect $T_c$ strongly depends on the detail of the electronic structure.
The problem in which materials or situations, the effect of the non-local terms becomes significant is highly non-trivial. 
%even in the case of simple $3d$ transition metal ferromagnets. 
We leave this interesting problem for future studies.

\section{Conclusion}
In this paper, we developed a local force method for the {\it ab initio} tight-binding model derived from wannierization of the SDFT Hamiltonian. In conventional Green's function formalism, spin-dependent hopping  (non-local magnetic potential) drastically increases the computational cost. To overcome this problem, we formulated a scheme based on KPM and performed a benchmark calculation for bcc-Fe, fcc-Co, and fcc-Ni. We found that the effect of spin-dependent hopping on $T_c$ is pronounced for bcc-Fe. We also presented several local approximations for spin-dependent hopping in the Green's function formalism, where the IRbasis significantly reduces the computational cost. We showed that approximation (A) in Sec. IV A works most successfully, in that it shows the best agreement with that of KPM. Our present approaches, which can be combined with any LSDA calculation regardless of the choice of the basis, would be an efficient scheme to evaluate $T_c$ of metallic magnets with a complex magnetic structure.

\section{Acknowledgement}
We are grateful to Y. Kato, A. Terasawa, H. Shinaoka, and T. Miyake for many valuable discussions. 
This work was supported by a Grant-in-Aid for Scientific Research (No.\ 19K14654, No.\ 19H05825, No.\ 19H00650, No.\ 18K03442, and No.\ 16H06345) from Ministry of Education, Culture, Sports, Science and Technology, and  CREST (JPMJCR18T3) from the Japan Science and Technology Agency.

\section*{Appendix}
\subsection{Derivation of some formulas in KPM}
In this appendix, we derive formulas of KPM-based approach given in the main text. First, we begin with Eqs.~\eqref{eq:kpm0} and \eqref{eq:alpha_} and derive the first derivatives \eqref{eq:kpm1} with Eq.~\eqref{eq:kpmbeta}. Let $\delta_x O$ denote the gradient of the given vector/matrix $O$ by the parameter $x$. From the definition of $\alpha_m$, we can easily see that $\delta_x\alpha_m$ can be expanded in terms of $(\delta_xA)\alpha_n$ ($0\le n \le m-1$) by the successive application of the chain rule. One may write this fact as the following form: 
\begin{align}
\delta_x\alpha_m = 2\sum_{n=0}^{m-1}P_{m,n}(\delta_xA)\alpha_n.\label{sup:1}
\end{align}
Here, the coefficient matrix $P_{m,n}$ for $n\ge1$ is given by $P_{m,m-1}=1$, $P_{m,m-2}=2A$, $P_{m,m-3}=4A^2-1$, $P_{m,m-4}=8A^3-4A$, and so on. The corresponding recursive relation is given by,
\begin{align}
P_{m,m-p}=2AP_{m-1,m-p}-P_{m-2,m-p},\label{sup:2}
\end{align}
for $m-1\ge p\ge2$ with $P_{m,m}=0$ and $P_{m,m-1}=1$. For $p=m$ component, we find $P_{m,0}=P_{m+1,1}/2$. By using Eq.~\eqref{sup:1}, we can expand $\delta_x F$ in terms of $(\delta_xA)\alpha_n$ as follows:
\begin{align}
\delta_x F&=2\sum_{m=0}^{M-1}\sum_{n=0}^{m-1} c_m r^\dagger  P_{m,n}  (\delta_x A)\alpha_{n}\\
&=2\sum_{n=0}^{M-2}\sum_{m=n+1}^{M-1}c_m r^\dagger P_{m,n}(\delta_xA)\alpha_n. \label{sup:4}
\end{align}
Namely, $\delta_x F=2\sum_{n=0}^{M-2}\beta_n (\delta_xA)\alpha_n$ where $\beta_n$ is given by,
\begin{align}
\beta_m&=\sum_{m=n+1}^{M-1}c_mr^\dagger P_{m,n}\\
&=c_{m+1}+2\beta_{m+1}A-\beta_{m+2}, \label{sup:3}
\end{align}
for $M-2\ge m\ge1$ and $\beta_m=0$ for $m\ge M-1$. Here, we have used Eq.~\eqref{sup:2}. Because of $P_{m,0}=P_{m+1,1}/2$, we have to divide Eq.~\eqref{sup:3} by two to obtain $\beta_0$. Finally, by replacing $x$ by $A_{12}$ and using $[(\delta_x A)\alpha_m]_{3}=\sum_{4}(\delta_{A_{12}} A_{34})[\alpha_m]_{4}=\delta_{13}[\alpha_m]_2$, we obtain Eq.~\eqref{eq:kpm1} with Eq.~\eqref{eq:kpmbeta} in the main text. 

Next, we derive Eq.~\eqref{eq:kpm2} with Eqs.~\eqref{eq:gamma1} and \eqref{eq:gamma2}.  Let us consider the the derivative of Eq.~\eqref{sup:4} by $y$:
\begin{align}
\delta_{xy}^2 F = 2\sum_{n=0}^{M-2}\left[(\delta_y\beta_n)(\delta_xA)\alpha_n+\beta_n (\delta_xA) (\delta_y\alpha_n)\right].\label{sup:5}
\end{align}
Here, we have used $\delta_{xy}A=0$ since we finally replace $x$ and $y$ by $A$.  For the second term of Eq.~\eqref{sup:5}, $\delta^2_{xy}F^{(2)}= 2\sum_{n=0}^{M-2}\beta_n (\delta_xA)(\delta_y\alpha_n)$, we can see,
\begin{align}
\delta^2_{xy}F^{(2)}
&=4\sum_{m=0}^{M-2}\sum_{n=0}^{m-1}\beta_m (\delta_xA) P_{m,n}(\delta_yA) \alpha_n\\
&=4\sum_{n=0}^{M-3}\sum_{m=n+1}^{M-2}\beta_m (\delta_xA) P_{m,n}(\delta_yA) \alpha_n\\
&=4\sum_{n=0}^{M-3}\tilde{\gamma}^{x}_n (\delta_yA)\alpha_n,
\end{align}
with the help of Eq.~\eqref{sup:1}. Then, the column vector $\tilde{\gamma}_n^x$ is evaluated by,
\begin{align}
\tilde{\gamma}_n^x&=\sum_{m=n+1}^{M-2}\beta_m (\delta_xA)P_{m,n}\\
&=\beta_{n+1}(\delta_xA)+2\tilde{\gamma}_{n+1}^x A-\tilde{\gamma}_{n+2}^x,
\end{align}
by using Eq.~\eqref{sup:2}. Similar to $\beta_n$, $\tilde{\gamma}_0^x$ is defined by = $\tilde{\gamma}_0^x= \frac{1}{2}(\beta_1(\delta_xA)+2\tilde{\gamma}_{1}^xA-\tilde{\gamma}_{2}^x)$ due to $P_{m,0}=P_{m+1,1}/2$.

For the first term, $\delta_{xy}^2F^{(1)}=2\sum_{n=0}^{M-3}(\delta_y\beta_n)(\delta_xA)\alpha_n$, first we expand $\delta_y\beta_n$ by $\beta_m(\delta_yA)$:
\begin{align}
\delta_y\beta_n = 2\sum_{m=n+1}^{M-2}\beta_m(\delta_yA) L_{m,n}.
\end{align}
Here, we used the fact $\beta_{m}=0$ for $m\ge M-1$. Now, we see that $L_{m,n}$ satisfies the following recurrence relation:
\begin{align}
L_{m,n}&=2AL_{m-1,n}-L_{m-2,n},
\end{align}
with $L_{n,n}=0$ and $L_{n+1,n}=1$. Based on these relations, we obtain,
\begin{align}
\delta_{xy^2}F^{(1)}&=4\sum_{n=0}^{M-3}\sum_{m=n+1}^{M-2}\beta_m(\delta_yA) L_{m,n}(\delta_xA)\alpha_n\\
&=4\sum_{m=1}^{M-2}\sum_{n=0}^{m-1}\beta_m(\delta_yA)L_{m,n}(\delta_xA)\alpha_n\\
&=4\sum_{m=1}^{M-2}\beta_m(\delta_yA) \gamma_m^x,
\end{align}
where $\gamma_m^x$ is given by,
\begin{align}
\gamma_m^x&=\sum_{n=0}^{m-1}L_{m,n}(\delta_xA)\alpha_n\\
&=(\delta_xA)\alpha_{m-1}+2A\gamma^x_{m-1}-\gamma^x_{m-2}.
\end{align}
Finally, by replacing $x$ by $A_{12}$ and $y$ by $A_{34}$, and using $[(\delta_yA)\alpha_n]_1=\delta_{31}[\alpha_n]^4$, and $[\beta_m(\delta_yA)]_{1}=\delta_{41}[\beta_m]^3$, we obtain Eq.~\eqref{eq:kpm2} with Eqs.~\eqref{eq:gamma1} and \eqref{eq:gamma2}.


\begin{thebibliography}{99}
\bibitem{memo01} Spin of the Wannier function is not identical to that of the spinor field in the presence of spin orbit coupling or non-collinear spin texture. However, we simply refer it spin when there is no confusion.
\bibitem{memo02} Note that $D(\bn_i,\theta_i)$ should be replaced by the combination of rotation matrices for spin and orbital representations in the presence of the spin-orbit coupling.
\bibitem{note03} Here, we neglect the correction coming from $|{\bm B}_{\rm eff}({\bm r})|^2$ term in the free energy since it does not contribute to the derivatives \eqref{eq:lich1} and \eqref{eq:lich2}.
\end{thebibliography}
\end{document}